\documentclass[pre,aps,twocolumn,superscriptaddress,nofootinbib]{revtex4}
\usepackage{dcolumn}
\usepackage{bm}
\usepackage{graphicx}
\usepackage{color}
\usepackage{subfigure}
\usepackage{hyperref}
\usepackage{latexsym}
\usepackage{amsthm}
\usepackage{amssymb}
\usepackage{array}
\usepackage{braket}
\usepackage{subfigure}
\usepackage{amsmath,amssymb}
\usepackage{bbm}
\usepackage{booktabs}
\usepackage{subfigure}
\usepackage{mathrsfs}
\usepackage{warpcol}
\usepackage{amsthm}

\usepackage{subfigure}

\usepackage{lipsum}
\usepackage{graphics}
\usepackage[dvipsnames]{xcolor}
\DeclareGraphicsExtensions{.jpg,.pdf, .mps, .png, .eps, .ps, .EPS,.gif}

\begin{document}
	\def\be{\begin{equation}}
	\def\ee{\end{equation}}
	
	\def\bc{\begin{center}}
		\def\ec{\end{center}}
	\def\bea{\begin{eqnarray}}
	\def\eea{\end{eqnarray}}
	\newcommand{\avg}[1]{\langle{#1}\rangle}
	\newcommand{\Avg}[1]{\left\langle{#1}\right\rangle}
	
	\def\ie{\textit{i.e.}}
	\def\etal{\textit{et al.}}
	\def\m{\vec{m}}
	\def\G{\mathcal{G}}

	\newcommand{\gin}[1]{{\bf\color{blue}#1}}
	\newcommand{\kir}[1]{{\bf\color{PineGreen}#1}}

	\newtheorem{theorem}{Theorem}
	\newtheorem{corollary}{Corollary}
	\newtheorem{lemma}{Lemma}
	\newtheorem{conjecture}{Conjecture}
	\newtheorem{proposition}{Proposition}
	\newcommand{\tr}{\ensuremath{\tilde{\bm\rho}}}
	\newcommand{\dsi}{\ensuremath{\Delta\bm\sigma_i}}
	
	\title{D-dimensional oscillators in simplicial structures: \\ odd and even dimensions display different synchronization scenarios}
	
	\author{X. Dai$^{*,}$}
	\affiliation{School of Mechanical Engineering, Northwestern Polytechnical University, Xi'an 710072, China}
	\affiliation{Center for OPTical IMagery Analysis and Learning (OPTIMAL), Northwestern Polytechnical University, Xi'an 710072, China}
	\affiliation{Unmanned Systems Research Institute, Northwestern Polytechnical University, Xi'an 710072, China}
	
	\author{K. Kovalenko$^{*,}$}
	\affiliation{Moscow Institute of Physics and Technology (National Research University), 9 Institutskiy per., Dolgoprudny, Moscow Region, 141701, Russian Federation}
	
	\author{M. Molodyk}
	\affiliation{Department of Physics, Stanford University, Stanford, CA 94305, USA}
	
	\author{Z. Wang}
	\affiliation{School of Mechanical Engineering, Northwestern Polytechnical University, Xi'an 710072, China}
	\affiliation{Center for OPTical IMagery Analysis and Learning (OPTIMAL), Northwestern Polytechnical University, Xi'an 710072, China}
	
	\author{X. Li$^{+,}$}
	\affiliation{Center for OPTical IMagery Analysis and Learning (OPTIMAL), Northwestern Polytechnical University, Xi'an 710072, China}
	
	\author{D. Musatov}
	\affiliation{Moscow Institute of Physics and Technology (National Research University), 9 Institutskiy per., Dolgoprudny, Moscow Region, 141701, Russian Federation}
	\affiliation{Russian Academy of National Economy and Public Administration, pr. Vernadskogo, 84, 119606 Moscow, Russia}
	\affiliation{Caucasus Mathematical Center at Adyghe State University, ul. Pervomaiskaya, 208, Maykop, 385000, Russia}
	
	\author{A. M. Raigorodskii}
	\affiliation{Moscow Institute of Physics and Technology (National Research University), 9 Institutskiy per., Dolgoprudny, Moscow Region, 141701, Russian Federation}
	\affiliation{Caucasus Mathematical Center at Adyghe State University, ul. Pervomaiskaya, 208, Maykop, 385000, Russia}
	\affiliation{Mechanics and Mathematics Faculty, Moscow State University, Leninskie Gory, 1, Moscow, 119991, Russia}
	\affiliation{Institute of Mathematics and Computer Science, Buryat State University, ul. Ranzhurova, 5, Ulan-Ude, 670000, Russia}
	
	\author{K. Alfaro-Bittner}
	\affiliation{Unmanned Systems Research Institute, Northwestern Polytechnical University, Xi'an 710072, China}
	\affiliation{Departamento de F\'isica, Universidad T\'ecnica Federico Santa Mar\'ia, Av. Espa\~na 1680, Casilla 110V, Valpara\'iso, Chile}
	
	\author{G. D. Cooper}
	\affiliation{Unmanned Systems Research Institute, Northwestern Polytechnical University, Xi'an 710072, China}
	
	\author{G. Bianconi}
	\affiliation{School of Mathematical Sciences, Queen Mary University of London, London, United Kingdom}
	\affiliation{The Alan Turing Institute, London, The British Library, United Kingdom}
	
	\author{S. Boccaletti}
	\affiliation{Unmanned Systems Research Institute, Northwestern Polytechnical University, Xi'an 710072, China}
	\affiliation{Moscow Institute of Physics and Technology (National Research University), 9 Institutskiy per., Dolgoprudny, Moscow Region, 141701, Russian Federation}
	\affiliation{CNR - Institute of Complex Systems, Via Madonna del Piano 10, I-50019 Sesto Fiorentino, Italy}
    \affiliation{Universidad Rey Juan Carlos, Calle Tulip\'an s/n, 28933   M\'ostoles, Madrid, Spain}

	\begin{abstract}
		From biology to social science, the functioning of a wide range of systems is the result of elementary interactions which involve more than two constituents, so that their description has unavoidably to go beyond simple pairwise-relationships. Simplicial complexes are therefore the mathematical objects providing a faithful
		representation of such systems. We here present a complete theory of synchronization of $D$-dimensional oscillators obeying an extended Kuramoto model, and interacting by means of 1- and 2- simplices. Not only our theory fully describes and unveils the intimate reasons and mechanisms for what was observed so far with pairwise interactions, but it also offers predictions for a series of rich and novel behaviors in simplicial structures, which include: a) a discontinuous de-synchronization transition at positive values of the coupling strength for all dimensions, b) an extra discontinuous transition at zero coupling for all odd dimensions, and c) the occurrence of partially synchronized states  at $D=2$ (and all odd $D$) even for negative values of the coupling strength, a feature which is inherently prohibited with pairwise-interactions.
		Furthermore, our theory untangles several aspects of the emergent behavior:  the system can never fully synchronize from disorder, and is characterized by an extreme multi-stability, in that the asymptotic stationary synchronized states depend always on the initial conditions.
		All our theoretical predictions are fully corroborated by extensive numerical simulations.  Our results elucidate the dramatic and novel effects that higher-order interactions may induce in the collective dynamics of ensembles of coupled $D$-dimensional oscillators, and can therefore be of value and interest for the understanding of many phenomena observed in nature, like for instance the swarming and/or flocking processes unfolding in three or more dimensions.
	\end{abstract}
	\maketitle
	
	$^{*}$ These Authors contributed equally to the Manuscript.
	
	$^{+}$ Corresponding Author: li@nwpu.edu.cn
	
	All collective properties emerging in complex systems arise from the specific way in which the elementary components interact \cite{vicsek_collective_2012,Herbert-Read18726,bricard_emergence_2013}.
	In past years, many relevant cases in physics, biology, social sciences and engineering have been successfully modelled as networks of coupled dynamical systems \cite{acebron2005kuramoto,strogatz2000kuramoto,physreport1,physreport2}.
	Such a representation, however, implies a too strong limitation, in that it explicitly assumes that the interplay among the system's units can always be factorized into the sum of pairwise interactions.
	Various recent studies have instead revealed that higher-order (many-body) interactions have to be accounted for a suitable representation of the structure and function of complex systems \cite{battiston2020networks,petri_simplicial_2018}. Examples include groups of actors in movies  \cite{ramasco_self-organization_2004}, spiking neuron populations  \cite{giusti_clique_2015, reimann_cliques_2017}, and co-authorship in scientific publications  \cite{patania_shape_2017}.
	Unravelling how such new types of interactions and their topology shapes the overall dynamics has thus been attracting  wide interest across disciplines \cite{parzanchevski_simplicial_2017,grilli_higher-order_2017,benson_simplicial_2018,bianconi2018topological,
		iacopini_simplicial_2019,millan2020explosive,torres2020simplicial}, and simplicial complexes are actually the proper mathematical objects to describe the structure of interactions among a complex system's units  \cite{salnikov_simplicial_2019,millan2019synchronization,millan2018complex,xu_bifurcation_2020,landry2020effect,st2020master,carletti2020random}. In the majority of cases, the interplay of the system's constituents leads to the raise of coordination: a phenomenon that is seen ubiquitously in biological and social contexts \cite{vicsek_collective_2012}. In particular, synchronous patterns are visible almost everywhere
	\cite{boccaletti_synchronization_2018}: from animal groups (bird flocks, fish schools, and insect swarms  \cite{vicsek_collective_2012,okeeffe_oscillators_2017}) to neurons in the brain \cite {singer_neuronal_1999}, and play a pivotal role in various functional aspects of real-world systems. Unveiling the essential mechanisms behind synchronization is therefore of great importance, and the Kuramoto model  \cite{araki_self-entrainment_1975} (with its various generalizations  \cite{strogatz2000kuramoto,acebron2005kuramoto,rodrigues2016kuramoto,boccaletti_explosive_2016}) is a fundamental reference for the study of such a phenomenon.
	
	In the original Kuramoto model, an ensemble of phase oscillators (each one rotating on the unit circle with a different natural frequency) is considered, with each oscillator being coupled with all the others through the sine function of the phase differences. In other words, the units are described via a single scalar variable (i.e., the phase on a 2-D circle).
	The model and its generalizations allowed to characterize a rich number of different dynamical states, from explosive synchronization to Chimera and Bellerophon states \cite{boccaletti_explosive_2016,ji2013cluster,laing2009dynamics,sethia2008clustered,abrams2004chimera,moreno_synchronization_2004,kuramoto_coexistence_2002, bi_coexistence_2016, martens_exact_2009,shima_rotating_2004}, and have been used to study a wide variety of problems in biology, physics and engineering  \cite{ermentrout1991adaptive,marvel2009invariant,motter2013spontaneous}.
	Recently, the model has been modified to capture synchronization phenomena occurring with many-body interactions \cite{skardal2019higher} and higher-order topological  synchronization \cite{millan2020explosive} displaying explosive synchronization transitions.
	Most of the  studies have so far concentrated on $2$-dimensional phase oscillators, whereas in some relevant circumstances (like, for instance, swarmalators moving in 3-dimensional spaces or the Heisenberg model in the sense of mean-field  \cite{zhu_synchronization_2013,okeeffe_oscillators_2017,chandra_continuous_2019}) it is crucial to embed the oscillators in higher dimensional spaces in which synchronization behavior emerges.
	An example is the motion of a swarm or a flock of moving agents, which implies organization in a 3-dimensional space  \cite{zhu_synchronization_2013,chandra_continuous_2019}.
	While there are a few recent advances in the study of higher dimensional Kuramoto models \cite{chandra_complexity_2019,chandra_observing_2019,kong2020scaling,markdahl_high-dimensional_2020,PRL_Dai_2020}, the dynamics of such models in the presence of a simplicial structure of interaction is still unknown.
	
	In this paper, we report a complete theory for $D$-dimensional Kuramoto oscillators ($D\ge2$) interplaying by means of simplicial structures.
	The theory allows us first to fully describe all previously reported results  \cite{chandra_complexity_2019,chandra_observing_2019,kong2020scaling,markdahl_high-dimensional_2020} on higher dimension synchronization, and to extend then to arbitrary dimension $D$ the results obtained for  $D=2$ synchronization on simplicial complexes  \cite{skardal_abrupt_2019}. In particular, our theory unveils the fundamental reasons and mechanisms for the observed difference in odd and even dimensions.
	Moreover, the theory allows us to make a series of novel predictions about previously unreported dynamical features, which are then (one by one) verified by us with extensive numerical simulations. Namely, we demonstrate that: {\it i)} the synchronization transition is discontinuous for positive coupling strength at any dimension $D$ (even for odd dimensions, the transition at zero coupling has a discontinuous character); {\it ii)} even for negative coupling, a simplicial complex structure of interactions may determine partial synchronization at $D=2$, which is a totally new phenomenon, inherently prohibited to occur in the original Kuramoto model   \cite{hong_kuramoto_2011};  {\it iii)} multi-stability exists at all dimensions $D$, and there exist infinitely many stable synchronized states.
	
	The Manuscript is organized as follows: in Section I we introduce the model equations for the dynamics of an ensemble of $D$-dimensional Kuramoto oscillators inter-playing on pure simplicial complexes with $n$-simplex interactions for $n=1,2$. In Section II, we report all the novel phenomenology induced by the presence of simplicial interactions, which even includes states which are inherently prohibited for pairwise interacting oscillators. Section III describes the exact and rigorous theory behind our observations. In Section IV, we derive instead a simplified (approximated) theory for 2-simplex interactions (i.e. interactions involving three nodes) which has the great advantage of giving rather precise predictions at already very low computational costs.
	Finally, Section V contains our conclusive discussions.
	
	\section{$D$-dimensional Kuramoto model on simplicial complexes}
	
\begin{figure}[t]
		\centering
		\includegraphics[scale=1.0]{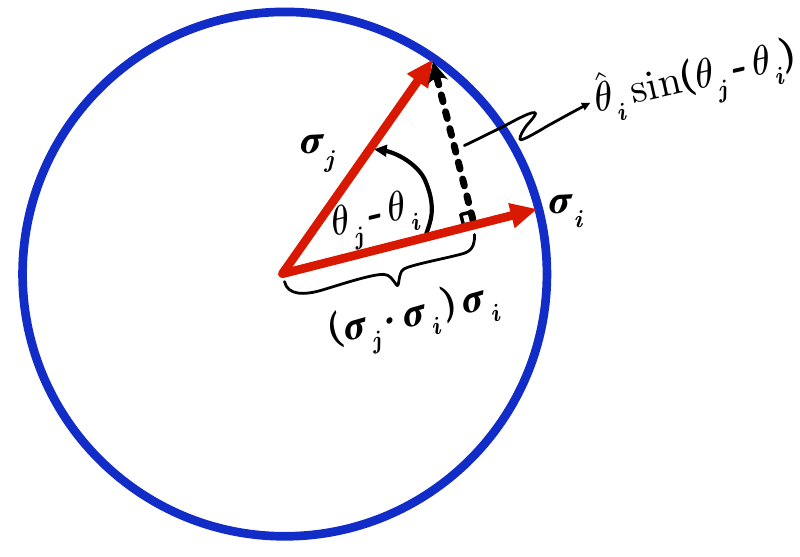}
		\caption{Schematic illustration of the vector representation of the Kuramoto model, at $D=2$. In the figure, $\bm\sigma_i$ and $\bm\sigma_j$ are two interacting oscillators.
			The main quantities appearing in Eqs.(3,4) are drawn in the Figure. $\hat{\theta}_i$ is a unit vector along the direction of the black dashed arrow.
		}
		\label{fig:0}
	\end{figure}

	Let us start by considering an ensemble of $N$ $D$-dimensional Kuramoto oscillators interplaying on pure simplicial complexes with $n$-simplex interactions (i.e. with interactions which involve $n+1$ oscillators).
	In such a system, each oscillator $i$ ($i=1,2\ldots, N$) is represented by a $D$ dimensional vector $\bm{\sigma}_i$ of norm one (i.e., a $D$ dimensional versor). Therefore, the trajectories of all $\bm\sigma_i$ are
	wandering on the hyper-surface of a $D$-dimensional unit sphere. Figure \ref{fig:0} is an illustration of the case $D=2$, which recovers the original Kuramoto model, where each oscillator $i$ is simply described by its phase $\theta_i$, i.e. $\bm\sigma_i=(\cos \theta_i,\sin \theta_i)$. Figure \ref{fig:0} contains also the explicit indication of all quantities that appear in the oscillators' evolution equations.
	
	In the absence of interactions, each vector rotates along some trajectory on the unit sphere dictated by a real anti-symmetric matrix $\bm{W}_i\in\mathbb{R}^{\text{D}\times D}$.  The matrices $\bm{W}_i$ are independently drawn at random for each node $i$. In particular, each upper triangular element of  $\bm{W}_i$ is sampled from a Gaussian distribution $\mathcal{N}(0,1)$, and the lower-triangular elements are fixed by the requirement that $\bm{W}_i$ must be anti-symmetric.
	For $D=2$, $\bm{W}_i\in\mathbb{R}^{2\times2}$ is given by
	\begin{equation}
	\bm{W}_i=
	\begin{pmatrix}
	0 & \omega_i \\
	-\omega_i & 0
	\end{pmatrix},
	\end{equation}
	where $\omega_i\sim \mathcal{N}(0,1)$ is the intrinsic frequency of node $i$.
	
	In other words, when the oscillators are independent from one another the equation ruling the dynamics of the ensemble is given by
	\bea
	\dot{\bm\sigma_i}=\bm{W}_i\bm\sigma_i.
	\label{eq:lambda=0}
	\eea
	
	When instead one considers a coupling among such oscillators (controlled by a non-zero coupling constant $\lambda\neq 0$), Ref.  \cite{chandra_continuous_2019}
	already shown that, as long as only 1-simplex interactions are taking place, the $D$-dimensional Kuramoto model can be reformulated as
	\bea
	\dot{\bm\sigma_i}&=&\bm{W}_i\bm\sigma_i+\frac{\lambda}{N} \sum_{j=1}^{N} [\bm\sigma_j-(\bm\sigma_j\cdot\bm\sigma_i)\bm\sigma_i].
	\label{equ:1simplex}
	\eea
	
	We here go beyond simple pairwise interactions, and focus on a $D$-dimensional Kuramoto model on 2-simplex structures, whose equations are given by
	\bea
	\dot{\bm\sigma_i}&=&\bm{W}_i\bm\sigma_i+\frac{\lambda}{2N^2}\left\{ \sum_{j=1}^{N}\sum_{k=1}^{N}(\bm\sigma_k\cdot\bm\sigma_i)[\bm\sigma_j-(\bm\sigma_j\cdot\bm\sigma_i)\bm\sigma_i]\right.\nonumber\\
	&&+\left.\sum_{j=1}^{N}\sum_{k=1}^{N}(\bm\sigma_j\cdot\bm\sigma_i)[\bm\sigma_k-(\bm\sigma_k\cdot\bm\sigma_i)\bm\sigma_i]\right\}.
	\label{equ:2}
	\eea
	
	Notice that, for $D=2$, our model fully encompasses the one treated in Ref. \cite{skardal_abrupt_2019}, and given by
	\begin{equation}
	\dot{\theta}_i=\omega_i+\dfrac{\lambda}{2N^2}\sum_{j=1}^{N}\sum_{k=1}^{N}\sin(\theta_j+\theta_k-2\theta_i).
	\label{equ:1}
	\end{equation}
	
	In order to monitor the level of coherence (synchronization) in the ensemble, we refer to a $D$-dimensional parameter given by
	\bea
	\bm\rho = \frac{1}{N}\sum_{i=1}^{N}\bm\sigma_i.
	\eea
	In all what follows, we will take as order parameter the quantity $R$ given by
	\bea
	R=|\bm\rho|,
	\eea
	where $|\ldots|$  indicates the $L_2$ norm.
	Additionally, we will indicate with $\hat{\bm \rho}=\bm\rho/R$ the unitary vector along the same direction of $\bm\rho$.
	Once again, our definition of $R$ is fully consistent with that used in the classical Kuramoto models, because for $D=2$  one has  $R=|\sum_{i=1}^{N}e^{\mathrm{i}\theta_i}/N|$.
	
	Eq.~(\ref{equ:2}) describes a system in which interactions occur between any triple of nodes. Moreover, it can be written in terms or the $D$-dimensional order parameter $\bm\rho$ as follows
	\begin{equation}
	\dot{\bm\sigma_i}=\bm{W}_i\bm\sigma_i+\lambda(\bm\rho\cdot\bm\sigma_i)[\bm\rho-(\bm\rho\cdot\bm\sigma_i)\bm\sigma_i].
	\label{equ:3}
	\end{equation}
	Similarly, Eq.~(\ref{equ:1simplex}) takes the form
	\bea
	\dot{\bm\sigma_i}=\bm{W}_i\bm\sigma_i+\lambda [\bm\rho-(\bm\rho\cdot\bm\sigma_i)\bm\sigma_i].
	\label{eq:1simplex}
	\eea
	The latter two Equations can be written in compact form as
	\bea
	\dot{\bm\sigma_i}=\bm{W}_i\bm\sigma_i+\lambda(\bm\rho\cdot\bm\sigma_i)^{n-1}[\bm\rho-(\bm\rho\cdot\bm\sigma_i)\bm\sigma_i],
	\label{equ:combined}
	\eea
	with $n\in \{1,2\}$ being the order of the simplices ($n=1$ for pairwise and $n=2$ for triadic interactions).
	Such $D$-dimensional Kuramoto models with 1-simplex and 2-simplex interactions are amenable to the self-consistent analytic treatment, as we will discuss in the following.
	In particular, our study focuses on the novel synchronization phenomena that are observed in $D$ dimensions (where complete synchronization means the perfect alignment of the
	$D$-dimensional vectors $\bm\sigma_i$) when one goes beyond pairwise-interactions.

	\section{The synchronization scenario}
	\subsection{The new phenomenology}
	
	The presence of simplicial interactions induces a remarkably rich phenomenology, which includes the rise of several
	novel states (i.e. dynamical states that are not observed in the original Kuramoto model).
	
	We focus on the desynchronization transition, i.e. the backward transition from the coherent state to incoherence. In our simulations, such transition can be monitored by initializing each oscillator $i$ (with probability $\mu$) along a unit $D$-dimensional vector $\bm{e}_D$, i.e.  $\bm\sigma_i=\bm{e}_D$. With probability $1-\mu$, the initialization is made instead in the opposite direction, i.e. $\bm\sigma_i=-\bm{e}_D$. After initialization of the vectors, one then starts simulating the system with a high value of the coupling strength $\lambda$ (a value for which the  synchronized state is stable) and progressively (and adiabatically) decreases $\lambda$ with steps $\delta \lambda=10^{-1}$.
	
	The latter statement implies that each value of $\lambda$ is kept constant for a lapse of time which is sufficient for the system to attain the new asymptotic state.
	Furthermore, in our simulations we use (unless otherwise stated) ensembles of  $N=5,000$ oscillators, and we integrate our ordinary differential equations by means of
	a fourth-order Runge-Kutta algorithm with integration time step $h=10^{-3}$.
	
\begin{figure}[t]
		\centering
		\includegraphics[width=\columnwidth]{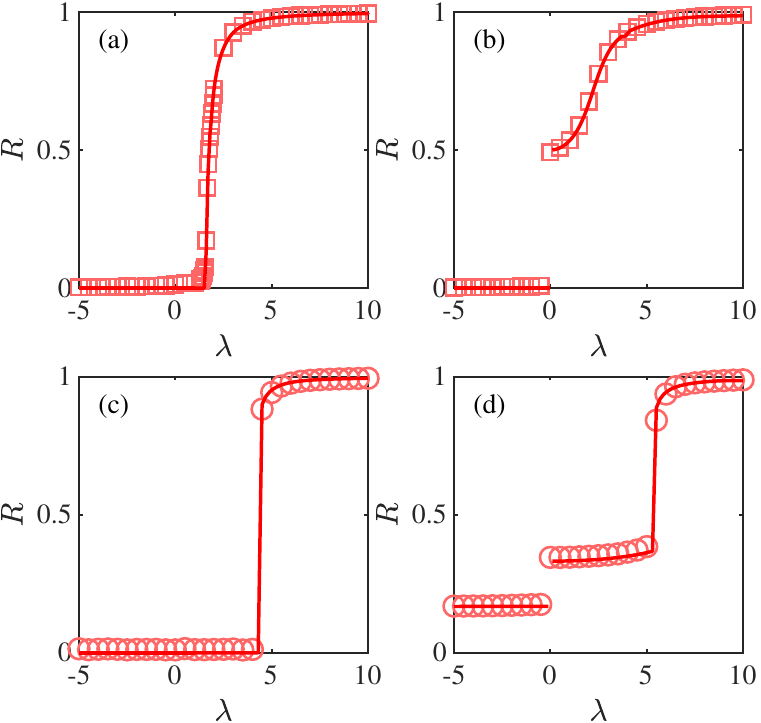}
		\caption{The Backward transition for (a) $1$-simplex interactions, $D=2$, (b) $1$-simplex interactions, $ D=3$, (c) $2$-simplex interactions, $D=2$, and (d) $2$-simplex interactions, $D=3$.
			$\mu=1$, and $\lambda$ is progressively decreased from $10$ to $0$. Subsequently, at $\lambda=0$, phases are re-initialized (setting again $\mu=1$) and $\lambda$ is gradually decreased from $0$ to $-5$. Red points indicate the simulation results, while red lines refer to the theoretical predictions obtained with the exact self-consistent approach, described in Sec. III.}
		\label{fig:1}
	\end{figure}

	Fig. \ref{fig:1}  reports the comparison of the desynchronization properties between the case of 1-simplex interactions (panels in the top row) and the case of 2-simplex interactions (panels in the bottom row), for $D=2$ (first column) and $D=3$ (second column).
	For networks with 1-simplex interactions, it is seen that the model supports a discontinuous transition at $\lambda=0$ for $D=3$ (top right panel), while for $D=2$ (the classical Kuramoto model) the transition is continuous and occurs at a non zero value of the coupling constant $\lambda$ (top left panel). Our results are fully consistent with those reported in Refs. \cite{chandra_complexity_2019,chandra_observing_2019}.
	In presence of 2-simplex interactions, one observes that the transition at $D=2$ becomes discontinuous (bottom left panel) and occurs at a non zero value of $\lambda$, $\lambda=\lambda_C>0$ (in  agreement with what recently reported in Ref. \cite{skardal_abrupt_2019}). For $D=3$ (bottom right panel) a rich and new phenomenology appears.  As one lowers the value of $\lambda$, the simplicial complex undergoes a discontinuous transition at a non zero value of $\lambda$, $\lambda=\lambda_C>0$ where the order parameter sharply decreases, but does not vanish. A partially coherent state is then set in the system which is robust against variations in the coupling parameter, up to reaching $\lambda=0$ (the uncoupled case) where the system features a second discontinuous transition toward another state with non-zero order parameter, which seems then to be stable also when $\lambda<0$. More precisely, if one just decreases $\lambda$ starting from 0, the order parameter will vanish, but partial coherence for negative $\lambda$ is seen if one re-initializes the phases of the system some at small negative $\lambda$.
	
\begin{figure}[ht]
		\centering
		\includegraphics[scale=1]{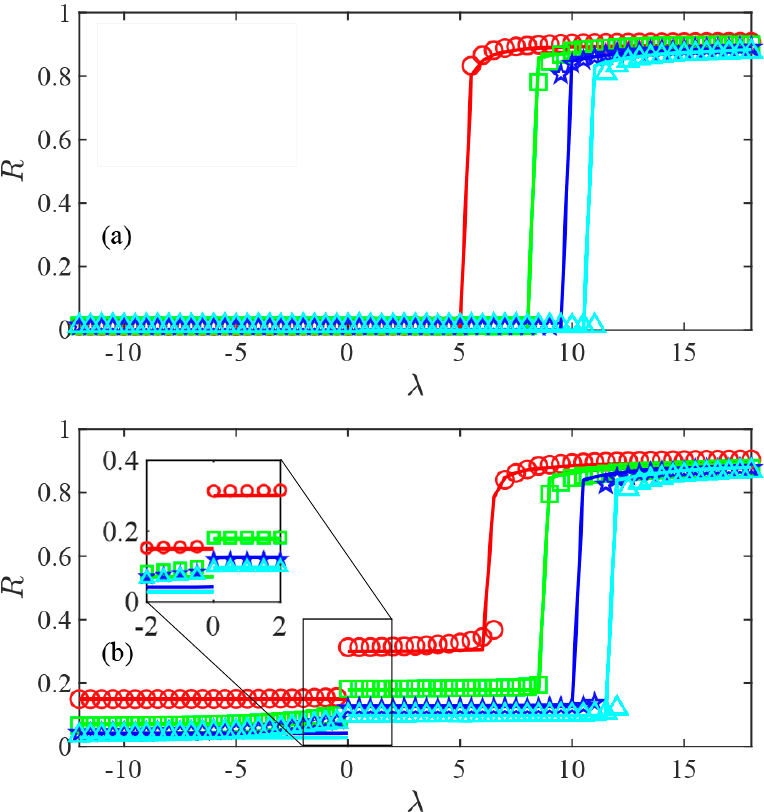}
		\caption{Order parameter $R$ (see text for definition) as a function of the coupling strength $\lambda$ for (a) even and (b) odd dimensions $D$. In all cases, points indicate the simulation results and solid lines refer to the theoretical predictions extracted by the approximated approach describe in Sec. IV.
			The simulation results are obtained by setting $\mu=0.95$, and gradually decreasing the value of $\lambda$ from $18$ to $0$. At $\lambda=0$, phases have been re-initialized (setting again $\mu=0.95$) and the scenario for $\lambda<0$ has been obtained by gradually decreasing $\lambda$ from $0$ to $-12$. Notice the common abrupt transitions at $\lambda>0$ in both panels, and the unique discontinuous transition at $\lambda=0$ for odd $D$ (which is magnified in the inset of panel b)). In panel a) the red, green, blue, and cyan curves and points correspond to $D=2,4,6$ and $8$ respectively. In panel b) the red, green, blue, and cyan curves and points correspond to $D=3,5,7$ and $9$ respectively.}
		\label{fig:2}
	\end{figure}

	In Fig. \ref{fig:2} we give further details on the observed desynchronization transitions in presence of 2-simplex interactions, and we report the backward transitions (stipulations are detailed in the Caption) for even [panel (a)] and odd [panel (b)] values of $D$, from $D=2$ to $D=9$.
	One can clearly see that the critical value $\lambda_C$ (i.e. the positive value of $\lambda$ at which the explosive desynchronization transition abruptly takes place) depends on the dimension $D$, but the qualitative features of the backward transition observed for $D$ even (for $D$ odd) are substantially reproducing those already discussed for $D=2$ (for $D=3$). In particular, looking at panel (b) of Fig. \ref{fig:2}, it is possible to see that the scenario described above of a double explosive transition from full coherence to partially coherent states is qualitatively conserved and robust at all odd dimensions.
	
	Finally, looking at Figs. \ref{fig:1} and \ref{fig:2}, the reader will certainly notice that all panels report colored points indicating the data coming from our simulations, but also solid lines which refer, instead, to the predictions offered by the exact and approximate treatments of our system, whose full details are the object of the next Sections and which actually are, in all cases, in remarkably good agreement with numerical simulations.

	\subsection{Self-consistent approach}
	
	Our theoretical analysis starts by considering a self-consistent approach, whose ultimate goal is to predict the stationary values of the order parameter $R$.
	To this end, let us consider a fixed and constant (i.e. time-independent) vector $\tilde{\bm\rho}$, and solve the equation
	\begin{equation}
	\label{eq:n-simplex_const}
	\dot{\bm\sigma_i} = \bm{W}_i \bm\sigma_i + \lambda(\tilde{\bm\rho}\cdot\bm\sigma_i)^{n-1} [\tilde{\bm\rho}-(\tilde{\bm\rho}\cdot\bm\sigma_i)\bm\sigma_i],
	\end{equation}
	for $n$-simplex interactions with $n=1,2$. In particular, one can find the stationary states $\bm\sigma_i^F$ of such a system of equations.
	
	Now, if the initial state is symmetric under the action of rotations around the $\tilde{\bm\rho}$ axis, one can assume that the real state of the system is also symmetric at any other time, and therefore the true value of the mean $\bm\rho$ will be collinear with $\tilde{\bm\rho}$.
	Such an important assumption allows one to calculate $R$ as the mean of the projections of the states $\bm{\sigma}_i^{F}$ onto the $\tilde{\bm\rho}$ axis. These latter projections (i.e. $\hat{\bm\rho} \cdot \bm\sigma_i^F$) will be, on their turn, functions of our choice of $\tr$ and of the random matrix $\bm{W}_i$.
	
	Therefore, one can evaluate the expected value $r(\tilde{R},\lambda)$ of the projection $\hat{\bm\rho} \cdot \bm\sigma_i^F$, given a fixed $\tilde{R}=|\tilde{\bm{\rho}}|$,  and a coupling constant $\lambda$ i.e.
	\bea
	\mathbb{E}(\hat{\bm\rho} \cdot \bm\sigma_i^F|\tilde{R})=r(\tilde{R},\lambda),
	\label{eq:12self}
	\eea
	and one immediately sees that $R$ must satisfy the following self-consistent equation
	\bea
	R=r(R,\lambda).
	\label{eq:self-R}
	\eea
	
	The solution(s) of Eq.~(\ref{eq:self-R}) as a function of $\lambda$ provide(s), in all cases, the theoretical predictions for the observed values of the order parameter.

	\subsection{Stability of the solutions}
	Once the stationary solutions $\bm\sigma_i^F$ of Eq.~$(\ref{eq:n-simplex_const})$ have been found, the next step is investigating their stability.
	By expanding close to $\bm{\sigma}_i^F$, one actually finds a significant difference between the cases of 1-simplex an 2-simplex interactions,
	which is at the basis of the new physical phenomena observed in Figs. \ref{fig:1} and \ref{fig:2}.
	
	Indeed, the linearized equation for 1-simplex interactions \cite{chandra_continuous_2019} reads as
	\bea
	\frac{1}{2}\frac{d}{dt}|\Delta \bm\sigma_i|^2 =
	- \lambda (\tilde{\bm\rho}\cdot\bm\sigma_i^F) |\Delta \bm\sigma_i|,
	\eea
	where $\Delta \bm\sigma_i=\bm{\sigma}_i-\bm\sigma_i^F$.
	It follows that the stationary solution is stable or unstable if the sign of $\lambda (\tilde{\bm\rho}\cdot\bm\sigma_i^F)$ is positive or negative, respectively.

	In 2-simplicial complexes, the linearized equation reads instead as
	\bea
	\hspace*{-5mm}\frac{1}{2}\frac{d}{dt}|\Delta \bm\sigma_i|^2 =
	\lambda |\Delta \bm\sigma_i|^2\left[\left(\tilde{\bm\rho}\cdot\frac{\Delta \bm\sigma_i}{\left|\Delta \bm\sigma_i\right|^2} \right)^2
	- \left(\tilde{\bm\rho}\cdot\bm\sigma_i^F\right)^2\right].\label{eq:stb_n2}
	\eea
	From this equation, one immediately infers that  $|\Delta \bm\sigma_i|^2$ changes exponentially slowly, and the stationary point $\bm\sigma_i^F$ is stable (unstable) if, on average, the factor
	\bea
	\lambda\left[\left(\tilde{\bm\rho}\cdot\frac{\Delta \bm\sigma_i}{\left|\Delta \bm\sigma_i\right|^2} \right)^2
	- \left(\tilde{\bm\rho}\cdot\bm\sigma_i^F\right)^2\right],
	\eea
	is negative (positive).

	In particular, for $D=2$ (i.e. in the classical Kuramoto setting) stability of the solution for $\lambda>0$ is warranted by the condition
	
	\bea
	\begin{array}{lcl}
		(\hat{\bm\rho}\cdot\bm\sigma_i^F)>\frac{1}{\sqrt{2}},
	\end{array}
	\eea
	while for $\lambda<0$ the solution is stable if and only if
	\bea
	\begin{array}{lcl}
		(\hat{\bm\rho}\cdot\bm\sigma_i^F)<\frac{1}{\sqrt{2}}.
	\end{array}
	\eea
	
	Note, indeed, that for small values of $\left|\dsi\right|$, the vector $\frac{\dsi}{\left|\dsi\right|}$ is nearly perpendicular to $\bm \sigma_i^F$ (since $\bm\sigma_i$ always remains on the unit sphere). Then, the vectors $\frac{\dsi}{\left|\dsi\right|}$ and $\bm\sigma_i^F$ can extended to a $D$-dimensional orthonormal basis, and one has
	
	\bea
	\begin{array}{lcl}
		\left(\hat{\bm\rho} \cdot \bm\sigma_i^F\right)^2 + \left(\hat{\bm\rho} \cdot \frac{\dsi}{\left|\dsi\right|}\right)^2 \leq 1,
	\end{array}
	\label{inequality}
	\eea
	as the left-hand side is approximately equal to the square of the projection of $\hat{\bm\rho}$ onto the plane spanned by $\frac{\dsi}{\left|\dsi\right|}$ and $\bm\sigma_i^F$.
	
	If $(\hat{\bm\rho}\cdot\bm\sigma_i^F)>\frac{1}{\sqrt{2}}$, then the solution is guaranteed to be stable for $\lambda>0$ and unstable for $\lambda<0$. Indeed, in this case
	\begin{equation}
	\left(\hat{\bm\rho} \cdot \frac{\dsi}{\left|\dsi\right|}\right)^2<\frac{1}{2}<\left(\hat{\bm\rho} \cdot \bm\sigma_i^F\right)^2,
	\end{equation}
	and therefore the factor $\lambda\left(\left(\tilde{\bm\rho}\cdot\frac{\dsi}{\left|\dsi\right|}\right)^2 - (\tilde{\bm\rho}\cdot\bm\sigma_i^F)^2\right)$ is always negative for $\lambda>0$ and always positive for $\lambda<0$.
	
	Moreover, in the two-dimensional case, the full basis consists of only two vectors, and the inequality Eq.~(\ref{inequality}) becomes an equality:
	
	\bea
	\begin{array}{lcl}
		\left(\hat{\bm\rho} \cdot \bm\sigma_i^F\right)^2 + \left(\hat{\bm\rho} \cdot \frac{\dsi}{\left|\dsi\right|}\right)^2 = 1.
	\end{array}
	\label{equality}
	\eea
	In such a case, if $(\hat{\bm\rho}\cdot\bm\sigma_i^F)<\frac{1}{\sqrt{2}}$, then
	\begin{equation}
	\left(\hat{\bm\rho} \cdot \bm\sigma_i^F\right)^2<\frac{1}{2}<\left(\hat{\bm\rho} \cdot \frac{\dsi}{\left|\dsi\right|}\right)^2,
	\end{equation}
	and therefore the factor $\lambda\left(\left(\tilde{\bm\rho}\cdot\frac{\dsi}{\left|\dsi\right|}\right)^2 - (\tilde{\bm\rho}\cdot\bm\sigma_i^F)^2\right)$ is always positive for $\lambda>0$ and always negative for $\lambda<0$.
	
	Notice that, for higher dimensions, the equality Eq.~(\ref{equality}) does not necessarily hold, and the value of $\left(\hat{\bm\rho} \cdot \bm\sigma_i^F\right)^2 + \left(\hat{\bm\rho} \cdot \frac{\dsi}{\left|\dsi\right|}\right)^2$ may change in time.
	
	For odd dimensions $D$, the system is in a synchronized state also for $\lambda\ll1$. By studying the stability of the solutions in this regime (see Appendix \ref{Ap:stability} for details), one finds some
	approximate criterion for the stability of the solution, expressed as
	
	\bea
	\begin{array}{lllcl}
		(\hat{\bm\rho}\cdot\bm\sigma_i^F)&>& Z_+(D) &\mbox{for} &\lambda>0, \\
		(\hat{\bm\rho}\cdot\bm\sigma_i^F)&<& Z_-(D) & \mbox{for} &\lambda<0.
	\end{array}
	\label{eq:stabDodd}
	\eea
	Here $Z_{\pm}(D)$ is given by
	\bea
	Z_\pm(D)= \sqrt{\frac{g_\pm(D)}{2+g_\pm(D)}},
	\eea
	where
	\bea
	g_{\pm}(D)=\int_{[0,+\infty)^{(D-1)/2}} d{\bf y} A_{\pm}({\bf y})\prod_{k=1}^{(D-1)/2}\left(\frac{1}{2}e^{-\frac{y_k}{2}}\right),
	\eea
	with
	\bea
	A_+({\bf y})&=&\left[\max_{1\leq k\leq(D-1)/2} y_k\right] \left[\sum_{k=1}^{(D-1)/2}y_k\right]^{-1},
	\\
	A_-({\bf y})&=&\left[\min_{1\leq k\leq(D-1)/2} y_k\right] \left[\sum_{k=1}^{(D-1)/2}y_k\right]^{-1},
	\eea
	and ${\bf y} \equiv (y_1, y_2, ...,y_{(D-1)/2})$ being a $[(D-1)/2]$-dimensional vector.

	\section{Exact approach}
	\subsubsection{General derivation}
	
	In the previous section we have discussed in wide generality how the $D$-dimensional synchronization of 1-simplex and $2$-simplex interaction can be derived using a self-consistent approach.
	
	Here we provide an exact approach for expressing the self-consistent Eq.~(\ref{eq:self-R}), where the function $r(\tilde{R},\lambda)$ is given by the expectation of the of projection of the vectors $\bm\sigma_i^F$ along $\tilde{\bm\rho}$  for a given value of $R=\tilde{R}$, as described by Eq.~(\ref{eq:12self}).
	
	In order to find the projection $\hat{\bm\rho} \cdot \bm\sigma_i^F$, we exploit the anti-symmetric nature of the matrices $\bm{W}_i$.
	Indeed, since $\bm{W}_i$ is antisymmetric, its eigenvalues are imaginary and come in conjugate pairs $\pm \mathbbm{i}\omega_i^{1},\cdots, \pm \mathbbm{i}\omega_i^{\lfloor D/2\rfloor}$, with an additional zero eigenvalue when $D$ is odd. Almost surely, $\omega^1_i,\cdots,\omega^k_i$ are all nonzero, so we will assume that this is the case. Without loss of generality, we can also take them all to be positive.
	
	For every node $i$, we can find an orthonormal basis in which $\bm{W}_i$ is a block diagonal matrix of the form
	\begin{equation}
	\label{eq:mat_even}
	{\bm W}_i=
	\begin{pmatrix}
	{\bm W}_i^{(1)} & 0 & \cdots & 0\\
	0 & {\bm W}_i^{(2)} & \cdots & 0\\
	\vdots & \vdots & \ddots & \vdots\\
	0 & 0 & \cdots & {\bm W}_i^{(D/2)}
	\end{pmatrix},
	\end{equation}
	if $D$ is even, and
	\begin{equation}
	\label{eq:mat_odd}
	{\bm W}_i=
	\begin{pmatrix}
	{\bm W}_i^{(1)} & 0 & \cdots & 0&0\\
	0 & {\bm W}_i^{(2)} & \cdots & 0&0\\
	\vdots & \vdots & \ddots & \vdots&\vdots\\
	0 & 0&\cdots & {\bm W}_i^{((D-1)/2)}&0\\
	0 & 0&\cdots & 0&0\\
	\end{pmatrix},
	\end{equation}
	if $D$ is odd, where we have indicated with ${\bm W}_i^{(k)}$ the $2\times 2$ matrix
	\begin{equation}
	{\bm W}_i^{(k)}=
	\begin{pmatrix}
	0 & \omega_i^{k}\\
	-\omega_i^{k} & 0
	\end{pmatrix}.
	\end{equation}
	In absence of interactions, for $\lambda=0$, the solution $\bm\sigma_i=\bm\sigma_i(t)$ to Eq.~(\ref{eq:lambda=0}) is therefore the result of  simultaneous rotations in several perpendicular planes with the angular velocities $\omega_i^{(1)},\cdots,\omega_i^{\lfloor D/2\rfloor}$. Then, if $D$ is even, $\bm\sigma_i$ does not have stationary points. On the other hand, if $D$ is odd, there are always two diametrically opposed stationary points of $\bm\sigma_i$. We will denote them $\pm\bm\sigma_i^{F,0}$.
	In the same basis one can represent $\bm{\sigma}_i$ and $\tilde{\bm \rho}$ as block vectors,
	\bea
	\label{eq:mat_even_2}
	\bm\sigma_i=
	\begin{pmatrix}
		\bm\sigma_i^{(1)} \\
		\bm{\sigma}_i^{(2)} \\
		\vdots\\
		\bm\sigma_i^{(D/2)}\\
	\end{pmatrix},& \tilde{\bm\rho}=
	\begin{pmatrix}
		\tilde{\bm\rho}^{(1)} \\
		\tilde{\bm{\rho}}^{(2)} \\
		\vdots\\
		\tilde{\bm\rho}^{(D/2)}\\
	\end{pmatrix},
	\eea
	for $D$ even, and
	
	\bea
	\label{eq:mat_odd_2}
	\bm\sigma_i=
	\begin{pmatrix}
		\bm\sigma_i^{(1)} \\
		\bm{\sigma}_i^{(2)} \\
		\vdots\\
		{\bm \sigma}_i^{\lfloor D/2\rfloor}\\
		\sigma_i^D\\
	\end{pmatrix}, &
	\tilde{\bm\rho}=
	\begin{pmatrix}
		\tilde{\bm\rho}^{(1)} \\
		\tilde{\bm{\rho}}^{(2)} \\
		\vdots\\
		\tilde{\bm\rho}^{\lfloor D/2\rfloor}\\
		\tilde{\rho}^{D}\\
	\end{pmatrix},
	\eea
	for $D$ odd where we have indicated with $\bm\sigma_i^{(k)}$ and $\tilde{\bm\rho}^{(k)}$ the vectors
	\bea
	\bm\sigma_i^{(k)}=
	\begin{pmatrix}
		\sigma_i^{2k-1}\\
		\sigma_i^{2k}
	\end{pmatrix}, &
	\tilde{\bm\rho}^{(k)}=
	\begin{pmatrix}
		\tilde{\rho}^{2k-1}\\
		\tilde{\rho}^{2k}
	\end{pmatrix}.
	\eea

	In presence of interactions for $\lambda\neq 0$ we search for stationary points of Eq.~(\ref{eq:n-simplex_const}) for all $\bm\sigma_i$ given a  fixed vector $\tilde{\bm\rho}$. For any given $i$ we consider the orthogonal basis such that $\bm W_i$ has the form as in Eq.~(\ref{eq:mat_even}). Recall that $\tilde{R}=|\tilde{\bm\rho}|$ and that  $\hat{\bm\rho}=\tilde{\bm \rho}/\tilde{R}$. Let us also write $\hat{\bm\rho}^{(k)}=\tilde{\bm \rho}^{(k)}/\tilde{R}$. By setting $\dot{\bm\sigma_i}=0$  in Eq.~(\ref{eq:n-simplex_const}) the following should hold for all $k=1,\cdots,\frac{D}{2}$:
	\begin{equation}
		\bm{W}_i^{(k)}\bm\sigma_i^{(k)}+\lambda \tilde{R}^n(\hat{\bm\rho}\cdot\bm\sigma_i)^{n-1}[\hat{\bm{\rho}}^{(k)}-(\hat{\bm\rho}\cdot\bm\sigma_i)\bm\sigma_i^{(k)}]={\bf 0}.
	\end{equation}
	Moreover if $D$ is odd we should also have
	\bea
	\lambda \tilde{R}^n(\hat{\bm\rho}\cdot\bm\sigma_i)^{n-1}[\hat{\rho}^{D}-(\hat{\bm\rho}\cdot\bm\sigma_i)\sigma_i^{D}]=0.
	\eea
	By rearranging these two equations one gets
	\bea
	\bm{W}_i^{(k)}\bm\sigma_i^{(k)}-\lambda \tilde{R}^n(\hat{\bm\rho}\cdot\bm\sigma_i)^n\bm\sigma_i^{(k)}=\lambda \tilde{R}^n(\hat{\bm\rho}\cdot\bm\sigma_i)^{n-1}\hat{\bm{\rho}}^{(k)}\nonumber\\
	\lambda \tilde{R}^n(\hat{\bm\rho}\cdot\bm\sigma_i)^{n}\sigma_i^{D}=\lambda \tilde{R}^n(\hat{\bm\rho}\cdot\bm\sigma_i)^{n-1}{\hat{\rho}}^{D}.
	\label{uno}
	\eea
	Let us now observe that $\bm{W}_i^{(k)}\bm\sigma_i^{(k)}$ can be written as
	\bea
	\bm{W}_i^{(k)}\bm\sigma_i^{(k)}=\omega_i^k\begin{pmatrix} \sigma_i^{2k} \\-\sigma_i^{2k-1}\\\end{pmatrix},
	\label{due}
	\eea
	therefore we have
	\bea
	\left[\bm{W}_i^{(k)}\bm\sigma_i^{(k)}\right]\cdot  {\bm\sigma}_i^{(k)}&=&0.
	\label{trentaydos}
	\eea
	By squaring the first of Eq.~(\ref{uno}) and using Eq.~(\ref{trentaydos}) one gets
	\bea
	\label{eq:a2+b2=c2-2-even}
	(\lambda \tilde{R}^n(\hat{\bm\rho}\cdot\bm\sigma_i)^n)^2
	\left| \bm\sigma_i^{(k)}\right|^2+
	\left| \bm{W}_i^{(k)}\bm\sigma_i^{(k)}
	\right|^2\nonumber \\
	=
	(\lambda \tilde{R}^n (\hat{\bm\rho}\cdot\bm\sigma_i)^{n-1})^2
	\left|{\hat{\bm\rho}}_i^{(k)}
	\right|^2.
	\eea
	By using the notation
	\bea
	r_k&=&\left|{\hat{\bm\rho}}_i^{(k)}
	\right|=\sqrt{(\hat{\rho}_{2k-1})^2+(\hat{\rho}_{2k})^2},\nonumber \\
	l_k&=&\left| \bm\sigma_i^{(k)}\right|=\sqrt{(\sigma_i^{2k-1})^2+(\sigma_i^{2k})^2},
	\label{eq:rk_lk}
	\eea
	and using Eq.~(\ref{due}) one finds the following expression for the square of $l_k$
	\begin{equation}
	(l_k)^2
	=
	\frac{(r_k)^2(\hat{\bm\rho}\cdot\bm\sigma_i)^{2(n-1)}}{(\hat{\bm\rho}\cdot\bm\sigma_i)^{2n}+\left(\frac{\omega_k^{(i)}}{\lambda \tilde{R}^n}\right)^2}.
	\label{eq:lk}
	\end{equation}
	By following similar steps starting from the second of Eq.~(\ref{uno}) for odd values of $D$ one gets
	\begin{equation}
	(l_D)^2=|\sigma_i^D|=\frac{\hat{\rho}_D^2}{(\hat{\bm\rho}\cdot\bm \sigma_i)^2},
	\label{eq:lD}
	\end{equation}
	as long as $\hat{\bm\rho}\cdot\bm \sigma_i\neq 0$.
	Since $\bm\sigma_i$ is a unitary vector one must have
	\bea
	|\bm\sigma_i|^2=\sum_{k=1}^{\lfloor D/2\rfloor}l_k^2+l_D^2\delta_{D,2\lfloor D/2\rfloor +1} =1,
	\eea
	where here and in the following $\lfloor x\rfloor$ indicates the floor function and $\delta_{x,y}$ indicates the Kronecker delta.
	
	By using Eqs.~(\ref{eq:lk})-(\ref{eq:lD}), this normalization condition applied to the stationary value $\bm\sigma_i^F$ reads
	\bea
	1&=&\sum_{k=1}^{\lfloor D/2\rfloor} \frac{(\hat{\rho}_{2k-1}^2+\hat{\rho}_{2k}^2)(\hat{\bm\rho}\cdot \bm\sigma_i^F)^{2(n-1)}}{(\hat{\bm\rho}\cdot \bm\sigma_i^F)^{2n}+\left(\frac{\omega_i^{k}}{\lambda \tilde{R}^n}\right)^2}\nonumber\\&&+\delta_{D,2\lfloor D/2\rfloor+1}\frac{\hat{\rho}_D^2}{(\hat{\bm\rho}\cdot\bm\sigma_i^F)^2},\label{eq:n_self}
	\eea
	{where for $D$ odd this equation is only valid for $\hat{\bm\rho}\cdot \bm\sigma_i^F\neq 0$.}
	
	This equation must be satisfied by any stationary state $\bm{\sigma}_i^F$ and can be interpreted as an equation for the projection of this state along the direction of $\tilde{\bm\rho}$, i.e. an equation for $\hat{\bm\rho}\cdot \bm\sigma_i^F$ given the choice of node $i$ internal frequencies $\bm \omega_i$ and the vector $\tilde{\bm \rho}$.

	For $n=1$, in presence of 1-simplex interactions, it can be shown that for odd $D$ (when $\hat{\rho}_D \neq 0$)  there is always exactly one solution for $|\hat{\bm\rho} \cdot {\bm\sigma}_i^F|$, while for even $D$ a solution only exists for large enough values of $\lambda$. Therefore if  Eq.~(\ref{eq:n_self}) has a non-zero solution it has   always has exactly one positive solution and one negative solution for $\hat{\bm\rho}\cdot \bm\sigma_i^F$. As already shown, for $\lambda>0$ the positive solution is stable and the negative one is unstable, while for $\lambda<0$ the opposite is true. It can also be shown that, for both $1-$ and $2-$simplices, the solution always satisfies $|\hat{\bm\rho}\cdot{\bm\sigma}_i^F |\geq \hat{\rho}_D$.

	For 2-simplex interactions, (i.e. for $n=2$) as for the case of 1-simplex interactions a solution is guaranteed to exist for odd $D$ (for $\hat{\rho}_D \neq 0$). However, regardless of the parity of $D$, in presence of 2-simplex interactions, there might be more than one solution for $|\hat{\bm\rho}\cdot {\bm\sigma}_i^F|$.
	
\begin{figure}[t]
		\centering
		\includegraphics[width=\columnwidth]{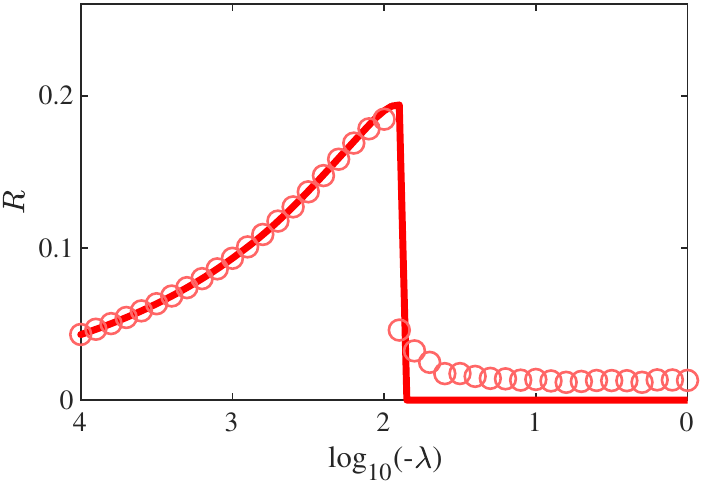}
		\caption{The order parameters $R$ vs. $\lambda<0$ for $D=2$, and 2-simplex interactions. The simulations results are obtained by setting an initial condition with $\mu=1$, and by starting with the value $\lambda=-10^4$.
			Subsequently, $\log_{10}(-\lambda)$ is gradually decreased with step $\delta= 0.1$.
			In the figure, red circles indicate the numerical results, and the solid lines indicate the theoretical predictions obtained with the approximated approach described in Sec. IV. Results are obtained on one network realization, and are the result of an average over the last $2\times10^4$ steps, with a total step of $2\times10^5$.
		}
		\label{fig:3}
	\end{figure}

	There are two important differences from the 1-simplex case. Firstly, there can now be unstable stationary points with $\hat{\bm\rho}\cdot\bm\sigma_i^F>0$ for $\lambda>0$. Secondly, stationary points whose projections onto the $\hat{\bm\rho}$ axis differ only by a sign have the same stability properties. If the sign of $\lambda$ is changed, stable points become unstable, and vice versa. This implies that, if there are unstable stationary points with $\hat{\bm\rho}\cdot{\bm\sigma}_i^F>0$ for positive $\lambda$, there can be coherent states even when $\lambda<0$ (unlike the 1-simplex case). This effect is observed in odd dimensions (see Fig. \ref{fig:1} and Fig.~\ref{fig:2}) and for $D=2$ (see Fig.~\ref{fig:3}). For even dimensions, our theory does not prove nor disprove the existence of such coherent states, but we have not observed them in our simulations.

	\subsection{1-simplex}

	For 1-simplex interactions the exact self-consistent Eq.~(\ref{eq:n_self}) has at most one positive root and one negative root, and for positive $\lambda$ the positive root corresponds to the stable stationary point while the negative root corresponds to the unstable stationary point. For negative $\lambda$ the opposite is true. When it exists, let us denote by $h(\bm \omega,\hat{\bm\rho},\tilde{r},\lambda)$
	the stable solution $\hat{\bm\rho}\cdot \bm\sigma_i^F$ of Eq.~(\ref{eq:n_self}), i.e.
	\bea
	h(\bm\omega,\hat{\bm\rho},\tilde{R},\lambda)=\left\{\begin{array}{ll}
		\hat{\bm\rho}\cdot \bm\sigma_i^F &\mbox{for}\ \  \hat{\bm\rho}\cdot \bm\sigma_i^F>0,\\
		0& \mbox{otherwise}. \end{array}\right.
	\eea
	With this notation the expression for $r(\tilde{R},\lambda)$ appearing in the self-consistent Eq.~(\ref{eq:self-R}) reads
	\bea
	r(\tilde{R},\lambda)=\int h(\bm\omega,\hat{\bm\rho},\tilde{R},\lambda) G(\bm{\omega}){ U}({\hat{\bm \rho}})d\bm\omega d\hat{\bm \rho},
	\eea
	where $G(\bm\omega)$ is the known distribution of eigenvalues of random antisymmetric hermitian matrices \cite{mehta1968distribution} and $U({\hat{\bm \rho}})$
denotes the uniform probability distribution of the vector $\hat{\bm\rho}$ on the (D-1)-dimensional unit sphere. For positive values of $\lambda$ this approach describes the $D$-dimensional Kuramoto synchronization transition (see Fig. $\ref{fig:1}$(a-b) for $\lambda>0$).
	For $\lambda<0$ we should take the negative root of the equation Eq.~(\ref{eq:n_self}), so
	\bea
	r(\tilde{R},-|\lambda|)=-r(\tilde{R},|\lambda|),
	\eea
	and the only solution of the self-consistence equation $R=r(R,-|\lambda|)$ is zero (see Fig. $\ref{fig:1}$(a-b) for $\lambda<0$).

	\subsection{2-simplex}
	
	As already mentioned in the preceding paragraphs for 2-simplex interactions Eq.~(\ref{eq:n_self}) might have more than two solutions.
	Higher-order interactions introduce significant numerical complications when we want to find the stable solution of Eq.~(\ref{eq:n_self}).
	In order to have an efficient algorithm to find the order parameter of the synchronization dynamics in the next section we will propose an approximated approach to derive the self-consistent equation of $R$.
	However, some important results can be obtained also with the present exact approach as we outline below.
	In presence of 2-simplex interactions, let us treat separately the case for $D$ odd and for $D$ even. Furthermore, here we assume that $\mu=1$, i.e., all agents are initialized at $\bm{e_D}$, while the dependence on $\mu$ will be discussed later on.

	For $D$ even and positive $\lambda$, let us indicate with $h(\bm \omega,\hat{\bm\rho},\tilde{R},\lambda)$
	the {\em maximum positive solution} $\hat{\bm\rho}\cdot \bm\sigma_i^F>0$ of Eq.~(\ref{eq:n_self})  (when it exists), and  $h(\bm \omega,\hat{\bm\rho},\tilde{R},\lambda)=0$ otherwise.
	Therefore the expression for $r(\tilde{R},\lambda)$ is given in this case by
	\bea
	r(\tilde{R},\lambda)=\int h(\bm \omega,\hat{\bm\rho},\tilde{R},\lambda) G({\bm\omega}){ U}(\hat{\bm\rho})d\bm\omega d\hat{\bm\rho}.
	\label{eq:rR}
	\eea
	Using this expression for $r(\tilde{R},\lambda)$ we can obtain excellent agreement with the simulation results (see Fig.~\ref{fig:1}(c)).
	The proliferation of many stationary points implies that stable stationary points with $\hat{\bm\rho}\cdot\bm\sigma_i^F>0$ may exist also for negative values of $\lambda$. Thus, coherent states can exist for $\lambda<0$. It should be remarked that this effect constitutes a fully novel feature of our model and is intimately related to the presence of 2-simplex interactions. Indeed, coherent states in the classical Kuramoto system (i.e. for $D=2$) are intrinsically prohibited at negative values of the coupling strength in the case of pairwise interactions. Fig. $\ref{fig:3}$ reports the synchronization transition at negative $\lambda$ for 2-simplex interactions and $D=2$, and the emergence of the predicted partially coherent state is clearly seen. The implications of this novel state are very interesting. For instance, it is well known that if all agents (oscillators) of a community behave as {\it contrarians} (i.e. they try to oppose to the mean field) then no coherent dynamics is possible for pairwise interactions \cite{hong_kuramoto_2011}. Our results show that higher-order interactions may instead generate coherence also among groups of contrarians.
	
	For odd values of $D$, Eq.~(\ref{eq:n_self}) can have multiple positive and multiple negative solutions among which we search for a stable one. Finding the stable solution may be, however, a hard computational task. Therefore in order to express the value of $r(\tilde{R},\lambda)$ defined  by Eq.~(\ref{eq:12self}) we combine the exact self-consistent approach with the approximate criterion for the stability of the solution defined in Eq.~(\ref{eq:stabDodd}).
	To this end, for positive (negative) $\lambda$ we define  $h(\bm \omega,\hat{\bm\rho},\tilde{R},\lambda)$ to  be the maximum (minimum) solution of the equation Eq.~(\ref{eq:n_self}) if this root is larger than $Z_+(D)$ (smaller than $Z_-(D)$), and $0$ otherwise. Note that while Eq.~(\ref{eq:stabDodd}) was derived for small values of $\lambda$ here we use it as a stability criterion valid for any value of $\lambda$.
	Using this expression for $h(\bm \omega,\hat{\bm\rho},\tilde{R},\lambda)$, one can derive $r(\tilde{R},\lambda)$ using Eq.~(\ref{eq:rR}).
	By solving the self-consistent Eq.~(\ref{eq:self-R}) one can obtain the dependence of the order parameter $R$ as a function of $\lambda$, and finds an excellent agreement with simulations (see for instance Fig. $\ref{fig:1}$(d)).
	
	However, this approach is very computationally demanding. For this reason in the following section we propose an approximate self-consistent approach that is very efficient to find the order parameter $R=R(\lambda)$ for 2-simplex interactions.

	\section{Approximate approach for 2-simplex interactions}
	\label{sec:app}
	In this section we provide an approximate expression of $r(\tilde{R},\lambda)$ which allows us to solve the self-consistent Eq.~(\ref{eq:self-R}) and find an approximate solution for $R=R(\lambda)$ for $n=2$, i.e. in presence of 2-simplex interactions.
	
	We denote by $\alpha$ the stationary value of the angle between $\bm\sigma_i$ [satisfying Eq.~(\ref{eq:n-simplex_const})]  and $\hat{\bm{\rho}}$, i.e.  the angle $\alpha$ defined as
	\bea
	\hat{\bm\rho}\cdot\bm\sigma_i^F=\cos(\alpha).
	\eea
	One has then that  $|\dot{\bm\sigma_i}|=0$, where $\dot{\bm\sigma_i}$ satisfies Eq.~(\ref{eq:n-simplex_const}). By defining $v=|\bm{W}_i\bm\sigma_i^F|$, one then obtains that the angle $\alpha$ needs to satisfy the following equation
	\bea
	\frac{1}{2}|\lambda| \tilde{R}^2\sin(2\alpha)=v.
	\label{eq:stat_angle}
	\eea
	The velocity $v=|\bm{W}_i\bm\sigma_i^F|$ is a function of both $\bm{W}_i$ and $\bm\sigma_i^F$. In the basis in which $\bm{W}_i$ is block-diagonal we have the following explicit expression for $v$,
	\bea
	v=|\bm{W}_i\bm\sigma_i|=\sqrt{\sum_{k=1}^{\lfloor D/2\rfloor} (\omega_i^k l_k)^2},
	\eea
	where $l_k$ are defined in Eq.~(\ref{eq:rk_lk}) for  an arbitrary fixed point $\bm\sigma_i^F$.
	To eliminate the complicated dependence on $l_k$, we approximate $v$  as
	\bea
	v=\sqrt{\mathbb{E}\left(\sum_{k=1}^{\lfloor D/2 \rfloor} (\omega_i^k l_k)^2\right)},
	\eea
	where the expectation is taken over $l_k$ on the unit sphere, so $v$ is still dependent on $\bm{W}_i$ through the eigenvalues $\bm\omega_i$. By considering the fact that the elements of $\bm{W}_i$ are random antisymmetric hermitian variables, one can calculate the distribution  $p(v)$  of $v$ (see Appendix \ref{subsection:distr} for the details of this derivation).
	Thus, by assuming in Eq.~(\ref{eq:stat_angle})  that $v$ is known, one can find the explicit expression for the angle $\alpha$.

	It should be noticed that Eq.~(\ref{eq:stat_angle}) can only be satisfied for $v<\frac{1}{2}|\lambda|\tilde{R}^2$. In this case, for a given $v$ there are two angles $\alpha_{+},\alpha_-\in \left[0, \frac{\pi}{2}\right]$ which solve Eq.~(\ref{eq:stat_angle}):
	\bea
	\alpha_+(v)&=&\frac{1}{2}\arcsin\left(\frac{2v}{|\lambda| \tilde{R}^2}\right),\\\alpha_-(v)&=&\frac{\pi}{2}-\frac{1}{2}\arcsin\left(\frac{2v}{|\lambda| \tilde{R}^2}\right).\eea
	These two solutions correspond to the following two projections of the stationary points $\bm\sigma_i^F$ along the $\tilde{\bm\rho}$ axis:
	\bea
	h_{\pm}(v,\tilde{R})=\cos(\alpha_{\pm}(v)),
	\eea
	with \bea
	h_{\pm}(v,\tilde{R})=\sqrt{\frac{1}{2}\left(1\pm\sqrt{1-\left(\frac{2v}{\lambda \tilde{R}^2}\right)^2}\right)},
	\eea
	where we note  that $h_-(v,\tilde{R})\leq \frac{1}{\sqrt{2}}\leq h_+(v,\tilde{R})$ is always satisfied.
	Let us now consider separately the case of even and odd dimensions $D$.
	
	\subsection{Even dimensions}
	
	For even dimensions $D$ and positive values of $\lambda$, the stationary point $\bm\sigma_i^F$ with $(\hat{\bm\rho}\cdot\bm\sigma_i^F)\geq\frac{1}{\sqrt{2}}$ is stable. Therefore, for $v<\frac{1}{2}\lambda\tilde{R}^2$, we assume that $\bm\sigma_i$ converges to a stationary point making an angle $\alpha_+(v)$ with the $\tilde{\bm\rho}$ axis. However for $v>\frac{1}{2}\lambda\tilde{R}^2$,  there is no stationary point, and the time average of $\bm\sigma_i$ is zero.
	
	It follows that the function $r(\tilde{R},\lambda)$ is given by
	\begin{equation} \label{eq:r(R)-even-1}
	r(\tilde{R},\lambda) = \int\limits_{0}^{\frac{\lambda \tilde{R}^2}{2}} h_+(v,\tilde{R}) p(v)dv.
	\end{equation}
	For $\lambda<0$, the behavior in an arbitrary even dimension is quite complicated, but the case of $D=2$ can be treated by a similar analysis. Indeed, for $D=2$ and $\lambda<0$, the stable stationary point $\bm\sigma_i^F$ satisfies $(\hat{\bm\rho}\cdot\bm\sigma_i^F)\leq\frac{1}{\sqrt{2}}$.  Thus,  for $v<\frac{1}{2}|\lambda|\tilde{R}^2$, $\bm\sigma_i$ one needs to choose the angle $\alpha_-(v)$, while for $v>\frac{1}{2}|\lambda|\tilde{R}^2$ there is no stationary point leading to the following expression for $r(\tilde{R},\lambda)$
	\begin{equation} \label{eq:r(R)-even-2}
	r(\tilde{R},\lambda) = \int\limits_{0}^{\frac{|\lambda| \tilde{R}^2}{2}} h_-(v,\tilde{R})p(v)dv.
	\end{equation}
	By using Eqs.~(\ref{eq:r(R)-even-1})-(\ref{eq:r(R)-even-2}) in Eq.~(\ref{eq:self-R}) one obtains the order parameter $R$ as a function of the coupling constant $\lambda$, i.e. $R=R(\lambda)$.
	The predictions of this approximate solution are in excellent agreement with extensive numerical simulations, as it appears in Figs. $\ref{fig:2}$ and $\ref{fig:3}$.

	\subsection{Odd dimensions}
	\label{sec:app-odd}
	
	Let us recall that in the absence of interactions, there are two stationary points $\pm\bm\sigma_i^{F,0}$. Let $\bm\sigma_i^{F,0}$ denote the point that is closer to $\hat{\bm\rho}$.
	
	For odd dimensions, let us express $\bm{\sigma}_i^F\cdot \hat{\bm\rho}$ as a function of both $v$ and $z$, where
	$z=z_i$ denotes the projection of $\bm\sigma_i^{F,0}$ onto $\tilde{\bm\rho}$, i.e. $z_i=( \hat{\bm\rho}\cdot \bm\sigma^{F,0}_i)$.
	In general, $z_i$ will be a function of the random elements of the matrices ${\bm W}_i$, therefore here we consider $z$ as a random variable with distribution $q(z)$ evaluated in Appendix $\ref{subsection:distr}$.
	
	In the following we estimate the value of $\hat{\bm\rho}\cdot \bm{\sigma}_i^F$ for positive and negative values of $\lambda$.
	In our calculations, we will make extensive use of the fact that
	$\hat{\bm\rho}\cdot \bm{\sigma}_i^F\in [z_i,1]$.
	We remark that the arguments that we will provide in the following are approximate, and the validity of the assumptions is then checked by comparing the numerical simulations with our predictions.
	
	First, we consider the case $\lambda>0$. If $v\leq \frac{\lambda \tilde{R}^2}{2}$, we consider two cases. If $\cos(\alpha_{+}(v))>z_i$, we assume that $\bm\sigma_i$ converges to a stationary point making an angle $\alpha_+(v)$ with the $\tilde{\bm\rho}$ axis. In fact it  can be shown that $h_+(v,\tilde{R})>Z_+(D)$ is always true, implying that  this stationary point is stable. Otherwise we assume that $\bm\sigma_i$ converges to a stationary point $\bm\sigma_i^F\sim{\bm\sigma}_i^{F,0}$, so its projection onto the $\tilde{\bm\rho}$ axis will be approximately equal to $z_i$.
	
	If $v>\frac{\lambda \tilde{R}^2}{2}$, we also consider two cases. If $z_i<Z_+(D)$,  this implies that ${\bm\sigma}_i$ does not converge to a stationary point, and the time average of its projection onto the $\tilde{\bm\rho}$ axis is zero. On the other hand, if $z_i>Z_+(D)$, we assume   that ${\bm\sigma}_i$ converges the  stationary point   in absence of the interaction, i.e. ${\bm\sigma}_i^{F}\sim{\bm\sigma}_i^{F,0}$ therefore  $\hat{\bm\rho}\cdot\bm\sigma_i^F\simeq z_i$.

	\begin{figure}[t]
		\centering
		\includegraphics[width=\linewidth]{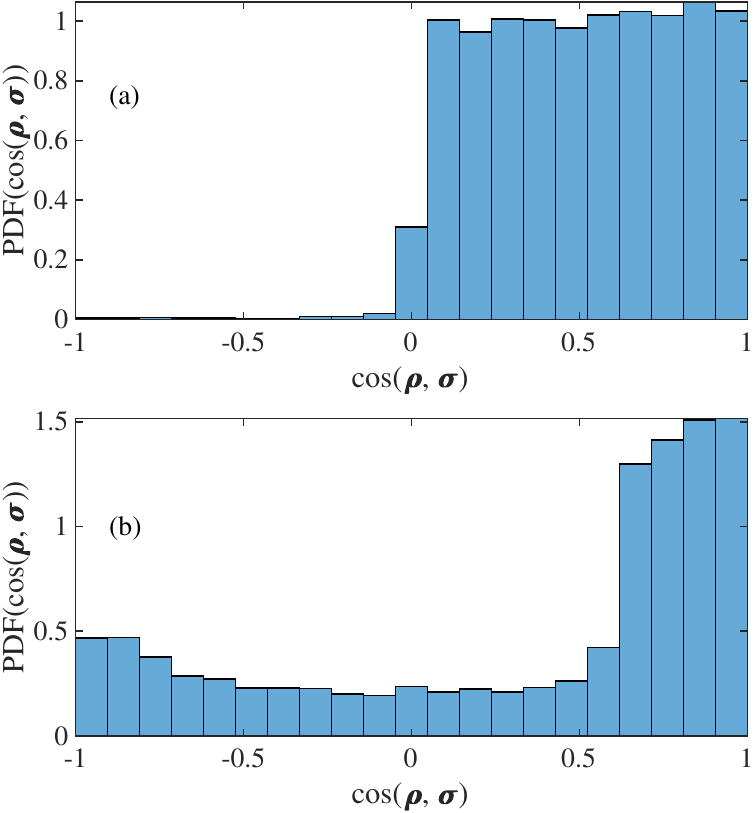}
		\caption{Histograms of the cosine function of the angles that the $N=5,000$ agents form (at $T=10^3$) with the system direction $\bm\rho$ for (a) 1-Simplex, and (b) 2-simplex with $D=3$. The agents are all initialized at $(1,0,0)$ and then evolved with a near-critical coupling strength $\lambda=0.5$. Nearly all these agents uniformly distribute on one hemisphere under pairwise-interactions (panel a), while agents in 2-simplex structures distribute more heterogeneously over the sphere (panel b).}
		\label{fig:4}
	\end{figure}
	
	\begin{figure}[t]
		\centering
		\includegraphics[scale=1]{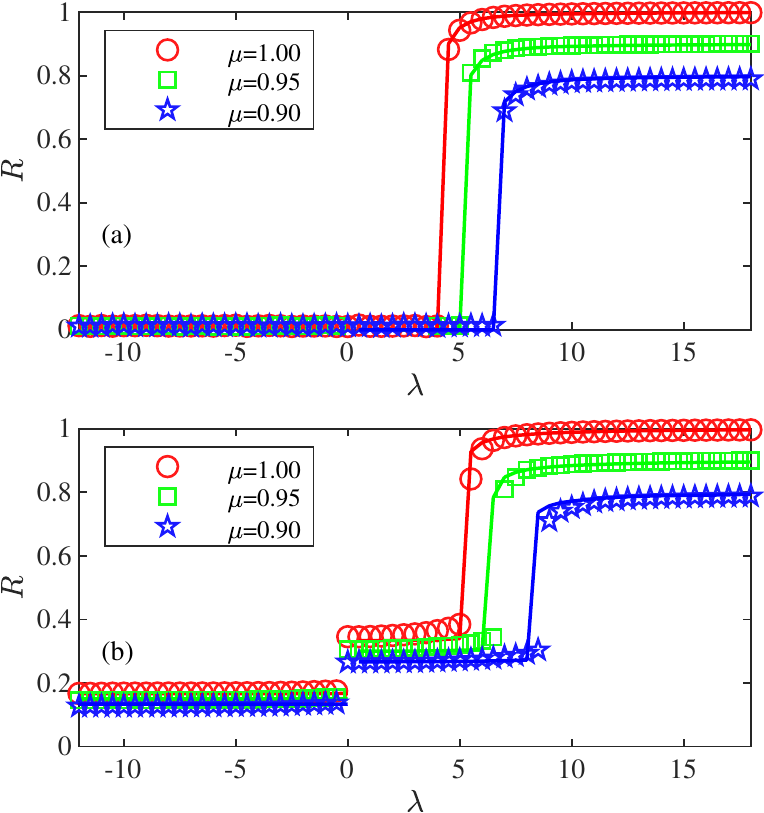}
		\caption{The origin of multi-stability in the backward synchronization transition for 2-simplex interactions. The order parameters $R$ as a function of the coupling strength $\lambda$  for (a) $D=2$, and (b) $D=3$. The legends report the color code for the curves reported in each panel, which correspond to different values of $\mu$ in the initial conditions. All backward transitions data are obtained with the same stipulations as in Fig. \ref{fig:2}. The solid lines are the theory predictions with the approximated approach.
			\label{fig:5}
		}
	\end{figure}

	The above arguments lead to the following expression for $r(\tilde{R},\lambda)$,
	\begin{equation} \label{eq:r(R)-odd-1}
	r(\tilde{R},\lambda)= \int\limits_{0}^{1}\int\limits_{0}^{+\infty} H_+(v,z,\tilde{R})p(v)q(z)dvdz,
	\end{equation}
	where $H_+(v,z,\tilde{R})$ is given by
	\bea
	H_+(v,z,\tilde{R}) =
	\begin{cases}
		M_+& 0\leq v\leq \frac{\lambda \tilde{R}^2}{2};\\
		z& Z_+(D)<z\text{ and } v > \frac{\lambda \tilde{R}^2}{2};\\
		0&z\leq Z_+(D) \text{ and } v > \frac{\lambda \tilde{R}^2}{2},
	\end{cases}
	\eea
	where $M_+=\max\left(h_+(v,\tilde{R}),z\right)$.

	Let us now consider the case $\lambda<0$. In this case, all stationary points with projections onto the $\tilde{\bm\rho}$ axis larger than $Z_-(D)$ are classified as unstable.
	
	In analogy to the $\lambda>0$ case, if $v\leq\frac{|\lambda|\tilde{R}^2}{2}$,  we take $\hat{\bm\rho}\cdot\bm\sigma_i^F=\max\left(h_-(v,\tilde{R}),z_i\right)$, as long as this expression is smaller than $Z_-(D)$. Otherwise, we take the time average of $\hat{\bm\rho}\cdot\bm\sigma_i^F$ equal to zero.
	
	If $v>\frac{|\lambda|\tilde{R}^2}{2}$, we assume that $\bm\sigma_i$ converges to a stationary point $\bm\sigma_i^F\sim\bm\sigma_i^{F,0}$ if $z_i\leq Z_-(D)$. Otherwise, there is no stationary point, and the time average of the projection of $\bm\sigma_i$ is zero. Accordingly,
	\begin{equation} \label{eq:r(R)-odd-2}
	r(\tilde{R},\lambda)= \int\limits_{0}^{1}\int\limits_{0}^{+\infty} H_-(v,z,\tilde{R})p(v)q(z)dvdz,
	\end{equation}
	where
	\bea
	H_-(v,z,\tilde{R}) =
	\begin{cases}
		M_- & M_-\leq Z_-(D)\text{ and }v\leq\frac{|\lambda|\tilde{R}^2}{2};\\
		z & z\leq Z_-(D)\text{ and } v>\frac{|\lambda|\tilde{R}^2}{2};\\
		0 & z> Z_-(D)\text{ and } v>\frac{|\lambda|\tilde{R}^2}{2};\\
		0 & M_-> Z_-(D)\text{ and }v\leq\frac{|\lambda|\tilde{R}^2}{2},
	\end{cases}
	\eea
	where $M_-=\max\left(h_-(v,\tilde{R}),z\right)$.
	By use of Eqs.~(\ref{eq:r(R)-odd-1}-\ref{eq:r(R)-odd-2}) in  Eq.~(\ref{eq:self-R}), one obtains $R=R(\lambda)$.
	In Fig. $\ref{fig:2}$ all our predictions are contrasted with extensive simulation results at different values of $D$ and the agreement is, once again, excellent.

	\subsection{Comparison between 1-simplex and 2-simplex interactions}
	
	One clear difference between the cases of 1-simplex and 2-simplex interactions is that the critical order parameter $R(\lambda\rightarrow0^+)$ is different in odd dimensions.
	For instance, for $D=3$, one has
	\begin{equation}
	R(\lambda\rightarrow0^+) = {1}/{3},
	\end{equation}
	which fits extremely well what one obtains with direct numerical simulations.
	
	The evidence of this difference stimulated us to perform further {\it microscopic} investigations, with the aim of unveiling the underneath reasons at the basis of such a crucially distinct behavior when $\lambda\rightarrow0^+$.
	
	Fig.~\ref{fig:4} contrasts the steady-state distributions of agents $\bm\sigma_i$ on the sphere in the cases of 1-simplex (see also Ref. \cite{chandra_continuous_2019}, where the parameter $K$ corresponds to $\lambda$ in our study) and 2-simplex interactions for $D=3$ at $\lambda=0.5$. In the 1-simplex case, as predicted, all agents converge to stable stationary points uniformly distributed over the hemisphere containing $\hat{\bm\rho}$. In the 2-simplex case, however, we see that there is a background distribution of agents roughly uniformly distributed over the sphere, due to agents that do not converge to a stationary point. The bins in which $\hat{\bm\rho}\cdot\bm\sigma_i>\frac{1}{\sqrt{3}}\approx 0.57$ contain many more agents due to agents converging to stable stationary points (with $\hat{\bm\rho}\cdot\bm\sigma_i^F>Z_+(3)=\frac{1}{\sqrt{3}}$), as predicted by our theory.
	It has to be remarked that, as long as we consider all-to-all connection structures, the 2-simplex model is actually equivalent to a 1-simplex model where the coupling $\lambda\left(\bm\rho\cdot\bm\sigma_i\right)$ is actually adaptive, and it is absolutely remarkable that high-order interactions alone can trigger such new kind of dynamics even when $\lambda\rightarrow0^+$.
	
	Another striking difference is the emergence of multistability in the 2-simplex case. 1-simplex interactions can be understood in terms of the oscillators experiencing an attraction towards their mean $\bm{\rho}$. In the case of 2-simplex interactions, however, they are attracted to the $\bm{\rho}$ \textit{axis} instead. Effectively, this means that an oscillator will be attracted to either $\bm{\rho}$ or $-\bm{\rho}$, whichever is closer. As was pointed out earlier, in the 2-simplex case, if $\bm\sigma_i^F$ is a stable stationary point of $\bm\sigma_i$, then so is $-\bm\sigma_i^F$. The consequence is that, in contrast to the 1-simplex case, agents can converge to stable stationary points in the hemisphere containing either $\bm\rho$ or $-\bm\rho$, depending on the initial conditions.
	
	If one initializes each $\bm\sigma_i$ at $\bm{e}_D$ with probability $\mu$ and at $-\bm{e}_D$ with probability $1-\mu$, then one can expect agents to converge to stationary points in the hemisphere in which they are initialized (with respect to $\bm{\rho}$). This affects our self-consistent analysis. Note that, in calculating $r(\tilde{R},\lambda)$, we always use the stationary points that are in same hemisphere as $\hat{\bm\rho}$, i.e. we assume $\mu = 1$. For $\mu<1$, we should assume that only a fraction $\mu$ of the converging oscillators will converge to stationary points in the same hemisphere as $\hat{\bm\rho}$, while the remaining $1-\mu$ converge to points in the opposite hemisphere. By symmetry, the oscillators in the opposite hemispheres will cancel each other out, and the value of $r(\tilde{R},\lambda)$ will be diminished by a factor of $2\mu-1$ compared to the case $\mu=1$, corresponding to the "surplus" of particles in the hemisphere containing $\bm{\rho}$. The self-consistence equation then becomes
	\begin{equation}
		R=(2\mu-1)r(R,\lambda).
	\end{equation}
	
	This results in the emergence of multistability, with a continuum of possible stationary values of the order parameter dependent on the initial conditions. Such a phenomenon was described in Ref. \cite{skardal_abrupt_2019} for a 2-simplex model in dimension $D=2$, and our analysis shows that it is actually present in the 2-simplex case for any dimension $D$. In Fig. \ref{fig:5}, multistability is illustrated for $D=2$ and $D=3$. The results of numerical simulations are in perfect agreement with self-consistent theory's predictions. Note that varying $\mu$ affects not only the stationary values of $R(\lambda)$, but also the values of $\lambda$ at which phase transitions occur. This is natural as the self-consistence equation is modified non-trivially.

	\section{Conclusions}
	
In summary, we presented a comprehensive framework to capture the effects that higher-order interactions have on the collective dynamics of coupled $D$-dimensional Kuramoto oscillators. 

Our theory and results allow to draw several conclusions and speculations.

First of all, it is demonstrated that the interplay between simplicial interactions and high dimensionality of the oscillators' dynamics gives rise to significant differences with respect to the standard $D=2$ dimensional Kuramoto model defined on graphs.
In particular, such higher-order interactions seem to prohibit synchronization transitions, and this may have relevance in neuroscience. Indeed, simplicial complexes have been largely observed in spiking neuron populations  \cite{giusti_clique_2015, reimann_cliques_2017}, and on the other hand it is well known that synchronization corresponds to brain's pathological states, like epilepsy. Therefore, the prevalence of simplicial interactions in such populations may occur because of the need to prevent local synchronization (and therefore malfunctioning) at a large scale of the brain activity, i.e. they can actually be imprinted on purpose in brain's architectures as local barriers to synchronous activity.

Furthermore, our results have demonstrated universality of the abrupt desynchronization transitions, and extensive multi-stability for any $D$-dimensional Kuramoto model with 2-simplex interactions.
At odd dimensions, the system displays moreover partial coherence even when the coupling strength vanishes (i.e. for $\lambda\rightarrow0^+$).
As a result, the system undergoes a two-stage abrupt desynchronization for odd dimensions. Also, the critical behaviors as coupling strength vanishes show dramatic difference for 1-simplex and 2-simplex, where all agents are distributed only on one hemisphere for the former.

Going ahead in our conclusions, we have revealed that the system can be found in a synchronized state also for negative couplings. This feature, fully induced by 2-simplex interactions, is a novel property of our model and may have implications, for instance, in social science. It is indeed well known that if all members (oscillators) of a community behave as {\it contrarians} (i.e. they are negatively coupled, and thus they try to oppose, to the mean field) then no coherent dynamics is possible for pairwise interactions \cite{hong_kuramoto_2011}. Our results show that higher-order interactions may instead generate partial coherence (or consensus) also among groups made of only {\it contrarians}, which is an interesting new phenomenon in social endeavours.

Finally, our study includes both an exact and an approximate theory for the prediction of the collective dynamics of coupled $D$-dimensional Kuramoto oscillators.
While our exact approach complements that already developed in Ref. \cite{chandra_continuous_2019} (and may find therefore applications in several other circumstances, as an alternative technique to analyze $D$-dimensional phase oscillators), the approximate approach furnished by us is totally new and may be of value in all those cases in which an exact treatment is fully or partially prevented, or when it would imply at some stage a too high computational demand.

Taken together, our work sheds therefore new light on the role that higher-order interactions have in determining the synchronization properties of coupled oscillators, opens new perspectives in the vibrant field of higher-order synchronization, and can hopefully stimulate further investigations (while indeed we solved here explicitly the case of an all-to-all configuration, one can certainly expect great differences and novel results when comparing networked topologies with structures where a given number of simplices -but not all possible ones- dictates the interactions).

\section{Acknowledgments}

We acknowledge support from
the National Natural Science Foundation of China (Grants No. U1803263, No. 11931015, and No. 81961138010),
Key Area R \& D Program of Shaanxi Province (Grant No. 2019ZDLGY17-07),
Key Area R \& D Program of Guangdong Province (Grant No. 2019B010137004),
the Fundamental Research Funds for the Central Universities (Grant No. 3102019PJ006),
the Russian Federation Government (Grant No. 075-15-2019-1926).
	

	\appendix
	\section{Description of stable and unstable points for $n=2$}
	\label{Ap:stability}
	In this Appendix we provide more details on the analytic derivation of Eq.~(\ref{eq:stabDodd}).
	From the linearized Eq.~(\ref{eq:stb_n2}) we see that, for a small perturbation $\dsi$ the norm $\left|\dsi\right|$ changes exponentially slowly. The stationary point $\bm\sigma_i^F$ is stable if,
	on average, the factor $\lambda\left(\left(\tilde{\bm\rho}\cdot\frac{\Delta \bm\sigma_i}{\left|\Delta \bm\sigma_i\right|}\right)^2 - (\tilde{\bm\rho}\cdot\bm\sigma_i^F)^2\right)$ is negative, and unstable if it is positive.
	
	However, performing this average over time for $D$ odd is challenging. Therefore, in order to derive a criterion for the stability of $\bm\sigma_i^F$, we adopt an approximate calculation of the average in the limit of small $\lambda$.
	We indeed observe that for small values of $\lambda$, the stationary solution $\bm\sigma_i^F$ in presence of the interactions is close to stationary solution $\bm\sigma_i^{F,0}$. Let us furthermore work in the orthonormal basis in which $\bm{W}_i$ is of the form shown in Eq.~(\ref{eq:mat_odd}). We suppose that $\bm\sigma_i$ stays close to the solution of $\dot{\bm\sigma_i}=\bm{W}_i\bm\sigma_i$ with fixed initial conditions (i.e., the influence of the term with $\tilde{\bm\rho}$ is small and
	\bea
	\bm\sigma_i^{(k)}=l_k\left(\begin{array}{c}\sin(\omega_i^{k} t+b_k)\\
		\cos(\omega^{k}_i t+b_k)\end{array}
	\right),
	\eea
	for some real $b_k$ and $l_k$ such that
	\bea
	\sum_{k=1}^{\frac{D-1}{2}}l_k^2=1-(\sigma_i^D)^2\approx |\dsi|^2,\eea
	and
	\bea\tilde{\rho}_D\approx \tilde{\bm\rho}\cdot\bm\sigma_i^F.\eea Consequently we can express $\left(\tilde{\rho}\cdot\frac{\dsi}{\left|\dsi\right|}\right)^2$ as
	\begin{widetext}
		\bea
		\left(\tilde{\bm\rho}\cdot\frac{\dsi}{\left|\dsi\right|}\right)^2 &=&
		\frac{1}{\sum_{k=1}^{\frac{D-1}{2}}l_k^2}
		\sum_{k=1}^{\frac{D-1}{2}}l_k^2\left((\tilde{\rho}_{2k-1})^2\sin^2(\omega^{k}_i t+b_k) +(\tilde{\rho}_{2k})^2\cos^2(\omega^{k}_i t+b_k)\right).
		\eea
	\end{widetext}
	The time average performed over   sufficiently long times is therefore given by
	\begin{equation}
		\left\langle\left(\tilde{\bm\rho}\cdot\frac{\dsi}{\left|\dsi\right|}\right)^2\right\rangle
		\approx
		\frac{1}{2}\frac{1}{\sum_{k=1}^{\frac{D-1}{2}}l_k^2}\sum_{k=1}^{\frac{D-1}{2}}l_k^2[(\tilde{\rho}_{2k-1})^2 +(\tilde{\rho}_{2k})^2].
	\end{equation}
	Let us now apply the following inequality to the right-hand side of the above equation:
	\bea
	\begin{split}
		\sum_{k=1}^{\frac{D-1}{2}}l_k^2[(\tilde{\rho}_{2k-1})^2 +(\tilde{\rho}_{2k})^2] \leq \\
		\leq \left(\sum_{k=1}^{\frac{D-1}{2}}l_k^2\right)\max\limits_{1\leq k \leq \frac{D-1}{2}}\left[(\tilde{\rho}_{2k-1})^2 +(\tilde{\rho}_{2k})^2\right].
	\end{split}
	\eea
	One needs the point to be stable for any possible value of $l_k$. Then, one obtains
	\bea
	\hspace*{-4mm}\left\langle\left(\tilde{\bm\rho}\cdot\frac{\dsi}{\left|\dsi\right|}\right)^2\right\rangle
	\approx \frac{1}{2}\max\limits_{1\leq k \leq \frac{D-1}{2}}\left[(\tilde{\rho}_{2k-1})^2 +(\tilde{\rho}_{2k})^2\right].
	\eea
	By imposing that
	\bea
	\lambda\left(\left\langle\left(\tilde{\bm\rho}\cdot\frac{\Delta \bm\sigma_i}{\left|\Delta \bm\sigma_i\right|}\right)^2\right\rangle - (\tilde{\bm\rho}\cdot\bm\sigma_i^F)^2\right)<0,
	\eea
	it follows that for $\lambda>0$ the stationary point of $\bm\sigma_i$ satisfies
	\bea
	\frac{1}{2}\max\limits_{1\leq k \leq \frac{D-1}{2}}\left[(\tilde{\rho}_{2k-1})^2 +(\tilde{\rho}_{2k})^2\right] < (\tilde{\bm\rho}\cdot\bm\sigma_i^F)^2.
	\label{eq:plus}
	\eea

	Analogously for $\lambda<0$ we have that the stationary point satisfies
	\bea
	\frac{1}{2}
	\min\limits_{1\leq k \leq \frac{D-1}{2}}\left[(\tilde{\rho}_{2k-1})^2 +(\tilde{\rho}_{2k})^2\right] < (\tilde{\bm\rho}\cdot\bm\sigma_i^F)^2.
	\label{eq:minus}
	\eea
	Recall that $\hat{\bm\rho} = \frac{\tilde{\bm\rho}}{\tilde{R}}$ and $\hat{\rho}_{i} = \frac{\tilde{\rho}_{i}}{\tilde{R}}$. To derive an approximate criterion depending only on $(\hat{\bm\rho}\cdot\bm\sigma_i^F)$, let us fix $(\hat{\bm\rho}\cdot\bm\sigma_i^F)$ and take the expectation value
	\bea
	\begin{split}
		\frac{1}{2} {\mathbb{E}} \left[\max\limits_{1\leq k \leq \frac{D-1}{2} }\left((\hat{\rho}_{2k-1})^2 +(\hat{\rho}_{2k})^2\right)\right]=\\
		=\frac{1}{2}(1-(\hat{\bm\rho}\cdot\bm\sigma_i^F)^2)g_+(D),\nonumber
	\end{split}
	\eea
	where
	\bea
	g_+(D)= \mathbb{E}\left[\max\limits_{1\leq k \leq \frac{D-1}{2}}((x_{2k-1})^2 +(x_{2k})^2)\right]
	\eea
	and $x=(x_1,\ldots,x_{D-1})$ is a vector uniformly distributed on the $(D-2)$-dimensional unit sphere.
	
	In this way, by defining $Z_{\pm}(D)$ as
	\bea
	Z_\pm(D)= \sqrt{\frac{g_\pm(D)}{2+g_\pm(D)}},
	\label{eq:ZP}
	\eea
	we obtain an approximate criterion for the stability of $\bm\sigma_i^F$ for $\lambda>0$, i.e.
		\begin{equation}
		(\hat{\bm\rho}\cdot\bm\sigma_i^F)> Z_+(D).
		\end{equation}
	Applying the same approximations starting from Eq.~(\ref{eq:minus}) we obtain that  for negative $\lambda$, $\bm\sigma_i^F$ is a stable stationary point if
	\begin{equation}
	(\hat{\bm\rho}\cdot\bm\sigma_i^F)< Z_-(D),
	\end{equation}
	where $Z_-(D)$ is defined in Eq.~(\ref{eq:ZP}).
	Note that $g_\pm(D)$ in Eq.~(\ref{eq:ZP}) can be expressed as
	\begin{equation}
		g_{\pm}(D)=\int_{[0,+\infty]^{(D-1)/2}} {d}{\bf y} A_{\pm}({\bf y})\prod_{k=1}^{(D-1)/2}\left(\frac{1}{2}e^{-\frac{y_k}{2}}\right),
	\end{equation}
	where  $A_\pm({\bf y})$ are defined as
	\bea
	A_+({\bf y})&=&\left[\max_{1\leq k\leq(D-1)/2} y_k\right] \left[\sum_{k=1}^{(D-1)/2}y_k\right]^{-1},
	\\
	A_-({\bf y})&=&\left[\min_{1\leq k\leq(D-1)/2} y_k\right] \left[\sum_{k=1}^{(D-1)/2}y_k\right]^{-1}.
	\eea

	\section{The distributions $p(v)$ and $q(z)$}
	\label{subsection:distr}
	In this Appendix we provide the derivation of the distributions $p(v)$ and $q(z)$ used in Sec. $\ref{sec:app-odd}$.
	
	To this purpose, let us recall that $v$ is defined as $v=|\bm{W}_i\bm\sigma_i|$,  and can be expressed as
	\bea
	v=\sqrt{\sum_{k=1}^{\lfloor D/2 \rfloor} \left(\omega_i^k l_k\right)^2}.
	\eea
	Here $\left(\omega_i^1,\ldots \omega_i^k\ldots \omega_i^{\lfloor D/2\rfloor}\right)$ are the angular velocities determined  by the random matrix $\bm{W}_i$ and $l_k$ are defined in Eq.~(\ref{eq:rk_lk}) which we rewrite here for convenience
	\bea
	l_k&=&\left| \bm\sigma_i^{(k)}\right|^2=\sqrt{(\sigma_i^{2k-1})^2+(\sigma_i^{2k})^2}.
	\label{eq:rk_lk2}
	\eea
	To simplify calculations, we estimate that, if $\bm\sigma_i$ has a stationary point, at that point $v$ can be approximated by the root-mean-square velocity given this $\bm{W}_i$:
	\bea
	v=\sqrt{\mathbb{E}\left(\sum_{k=1}^{\lfloor D/2 \rfloor}\left(\omega_i^k l_k\right)^2\right)},
	\eea
	where $\bm\sigma_i=(\sigma_i^1,\ldots,\sigma_i^D)$ is uniformly distributed on the unit sphere. Since $(\sigma_i^r)^2$ has the same distribution for all $r=1\ldots,D$ and $\sum_{r=1}^D (\sigma_i^j)^2=1$, for every $j$ we have $\mathbb{E}(\sigma_i^r)^2=\frac{1}{D}$. Using linearity of expectation, we find
	\bea\mathbb{E}\left(\sum_{k=1}^{\lfloor D/2 \rfloor} \left(\omega_i^kl_k\right)^2\right) =\frac{2}{D}\sum_{k=1}^{\lfloor D/2 \rfloor} \left(\omega_i^k\right)^2.
	\eea
	Therefore, in this approximation we obtain
	\bea
	v= \sqrt{\frac{2}{D}\sum_{k=1}^{\lfloor D/2 \rfloor}(\omega_i^k)^2}.
	\eea
	It follows that $v$ depends on the elements of $\bm{W}_i$. Since the matrix $\bm{W}_i$ is random we  treat $v$  as a random variable as well and we derive   its distribution $p(v)$.
	
	The non-zero eigenvalues of the matrix $\bm{W}_i$ are given by  $(\pm \mathbbm{i}\omega_i,\ldots,\pm \mathbbm{i}\omega_{\lfloor D/2\rfloor})$ where $\mathbbm{i}$ indicates the imaginary unit. Since for any matrix ${\bf A}$, the sum of the squares of the eigenvalues of ${\bf A}$ equals the trace of ${\bf A}^2$ we obtain
	\bea
	-2\sum_{k=1}^{\lfloor D/2\rfloor} \left(\omega_i^k\right)^2=\text{Trace}\left(\bm{W}_i^2\right).\eea
	At the same time, using the fact that $\bm{W}_i$ is antisymmetric, we find that
	\bea
	\hspace*{-3mm}\text{Trace}(\bm{W}_i^2) &=& \sum_{k=1}^{D}\sum_{l=1}^{D}w_{i}^{k,l}w_{i}^{l,k}=-2\sum_{k<l} (w_i^{k,l})^2,
	\eea
	where $w_i^{k,l}$ denotes the $(k,l)$ entry of $\bm{W}_i$ in the original basis. Thus,
	$$\sum_{k=1}^{\lfloor D/2\rfloor}(\omega_i^k)^2 =\sum_{k<l}(w_{i}^{k,l})^2.$$
	Since the $w_i^{k,l}$ are independent standard normal random variables, this means that $\sum_{k=1}^{\lfloor D/2\rfloor}(\omega_i^k)^2$ follows a $\chi\left(\frac{D(D-1)}{2}\right)$ distribution.
	Then the probability density function of $v$ has the following form:
	\begin{widetext}
		\begin{equation}
		\label{eq:distr_p(v)}
		p(v) = \frac{1}{2^{\left(\frac{D(D-1)}{4} - 1\right)} \Gamma\left(\frac{D(D - 1)}{4}\right) c(D)} \left(\frac{v}{c(D)}\right)^{\frac{D(D-1)}{2}-1} \exp\left(-\frac{1}{2}\left(\frac{v}{c(D)}\right)^2\right),
		\end{equation}
	\end{widetext}
	where $c(D) = \sqrt{\frac{2}{D}}$.
	
	We now turn to the random variable $z=\hat{\bm\rho}\cdot \bm\sigma^{F,0}_i $, which is a function of the stationary state ${\bm\sigma^{F,0}_i}$ of Eq.~(\ref{eq:lambda=0}) and ultimately a function of $\bm{W}_i$ . For $D\geq 3$, let $(x_1,\ldots,x_D)$ be a random vector uniformly distributed on the $(D-1)$-dimensional unit sphere. From the invariance of the distribution of $\bm{W}_i$ with respect to orthogonal change of basis, it follows that $z$ has the same distribution as $|x_k|$ for any fixed $k$. Therefore, it follows that the distribution $q(z)$ can be expressed as
	\begin{equation}
	\label{eq:distr_q(z)}
	q(z) =
	\begin{cases}
	\frac{2\Gamma(\frac{D}{2})}{\sqrt{\pi}\Gamma(\frac{D-1}{2})} (1-z^2)^{\frac{D-3}{2}} & 0\leq z \leq 1,\\
	0 & \text{otherwise.}
	\end{cases}
	\end{equation}
	
\end{document}